\documentstyle[prl,twocolumn,aps,amsfonts,psfig]{revtex}

\begin{document}

\twocolumn[\hsize\textwidth\columnwidth\hsize\csname @twocolumnfalse\endcsname

\title{Flip dynamics in octagonal rhombus tiling sets}

\author{{\sc N. Destainville}}
\address{Laboratoire de Physique Quantique~-- UMR CNRS-UPS 5626~-- IRSAMC,
Universit\'e Paul Sabatier, \\
118, route de Narbonne, 31062 Toulouse Cedex 04, France.}

\date{\today}

\maketitle

\begin{abstract}
We investigate the properties of classical single flip dynamics in
sets of two-dimensional random rhombus tilings. Single flips are local
moves involving 3 tiles which sample the tiling sets {\em via} Monte
Carlo Markov chains. We determine the ergodic times of these dynamical
systems (at infinite temperature): they grow with the system size
$N_T$ like $Cst.\; N_T^2 \ln N_T$; these dynamics are rapidly
mixing. We use an inherent symmetry of tiling sets and a powerful tool
from probability theory, the coupling technique. We also point out the
interesting occurrence of Gumbel distributions.

\medskip

\end{abstract}

\bigskip

]

\begin{figure}[ht]
\begin{center}
\ \psfig{figure=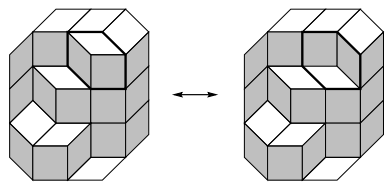,width=5cm} \
\end{center}
\caption{Examples of octagonal fixed boundary tiling and of elementary
flip. We have also displayed (in gray) the de Bruijn lines of a
family, among 4 families. They are lines of adjacent tiles sharing an
edge with a fixed orientation.}
\label{flip2D}
\end{figure}

After the discovery of quasicrystals~\cite{Shechtman84}, quasiperiodic
tilings~\cite{Penrose74} as well as their randomized counterpart,
random rhombus tilings~\cite{Henley91}, rapidly appeared to be
suitable paradigmatic models for quasicrystalline
alloys~\cite{Levine84}. Simultaneously, these systems also became an
active topic in discrete mathematics (see~\cite{Cohn98}
or~\cite{Latapy99} for examples).  Fig.~\ref{flip2D} displays a random
tiling, which belongs to the class of plane tilings with octagonal
symmetry. Beyond this case, plane tilings with larger symmetries
including Penrose tilings~\cite{Penrose74} and space tilings with
icosahedral symmetries were proposed to model every kind of
quasicrystal. The present letter is devoted to dynamical properties of
random rhombus tilings in terms of local dynamical rules, the
so-called {\em phason-flips}, which consist of local rearrangement of
tiles (Fig.~\ref{flip2D}). These dynamics are of interest for several
reasons. On the one hand, it is more and more clear that phason-flips
exist in real quasicrystals~\cite{Lyonnard96} and can be modeled in a
first approximation by tile-flips; they are a new source of atomic
mobility, as compared to usual crystalline materials. In particular,
they could carry their own contribution to
self-diffusion~\cite{Kalugin93}, even if the efficiency of such
processes remains controversial~\cite{Bluher98}. They are also
involved in some specific mechanical properties of quasicrystals, such
as plasticity {\em via} dislocation
mobility~\cite{Caillard99}. Therefore a complete understanding of flip
dynamics is essential in quasicrystal physics. The present work is a
first step in this direction. On the other hand, a lot of numerical
work has been carried out to characterize statistical properties of
tiling sets, a part of which was based on Monte Carlo techniques which
rely on a faithful sampling of tiling sets
(see~\cite{Tang90,Widom98}). So far, no systematic study of the
relaxation times between two independent numerical measures has been
accomplished, whereas it is an essential ingredient for a suitable
control of error bars. However, there exist exact results in the
simplest case of random rhombus tilings with hexagonal
symmetry~\cite{Randall,Randall98,Wilson99} and several estimates of
relaxation times in larger symmetries, either numerical or in the
approximate frame of Langevin dynamics~\cite{Tang90,Henley97}.

Random rhombus tilings are made of rhombi of unitary side length. They
are classified according to their global
symmetries~\cite{Henley91}. The simplest class of hexagonal tilings~--
made of 60$^{\rm o}$ rhombi with 3 possible orientations~-- has been
widely explored~\cite{Randall,Randall98,Wilson99}. Tilings with
octagonal symmetry are made of 6 different tiles (characterized by
their shape {\em and} orientation): two squares and four 45$^{\rm o}$
rhombi (Fig.~\ref{flip2D}). Beyond these two cases, one can define
tilings with higher symmetries ({\em e.g.}  Penrose
tilings~\cite{Penrose74}) or of higher dimensions~\cite{Henley91}.
For sake of technical simplicity, we focus on tilings filling a
centrally symmetric polygon with integral side lengths
(Fig.~\ref{flip2D}). We are interested in the large size limit where
the polygon becomes large keeping a fixed shape. Such tilings can
schematically be seen as frozen near their boundary, and strain-free
in their central region~\cite{Bibi98}. Dynamics of fixed
boundary tilings should be the manifestation of dynamics of their
strain-free center, thus relating both boundary conditions. Here we
suppose that all tilings have the same probability; we work at
infinite temperature.

The set of all the tilings of such a region together with the flip
dynamical rule define a discrete time Markov chain: at each step, a
vertex of the tiling is uniformly chosen at random and if this vertex
is surrounded by 3 tiles in flippable configuration, then we flip
it. Since sets of plane rhombus tilings are connected {\em via}
elementary flips~\cite{Bibi99}, this process can reach any tiling. It
converges toward the uniform equilibrium distribution, since it
satisfies detailed balance. All the difficulty is to characterize how
many flips one needs to get close to equilibrium.  Generally speaking,
let us consider a Markov chain on a finite configuration space $L$,
which converges toward a stationary distribution $\pi$. Let $x_0$ be
any initial configuration and $P(x,t|x_0,0)$ be the probability that
the process has reached the configuration $x$ after $t$ steps. Then
\begin{equation}
\Delta(t,x_0) = 1/2 \; \sum_{x \in L} | P(x,t|x_0,0) - \pi(x) |
\label{Delta0}
\end{equation}
usually measures the distance between both
distributions~\cite{Aldous81}. Given $\varepsilon >0$ we define the
{\em ergodic} or {\em mixing} time $\tau(\varepsilon)$ so that whatever $x_0$, after $\tau(\varepsilon)$ steps, one is sure
to stay within distance $\varepsilon$ of equilibrium:
\begin{equation}
\tau(\varepsilon) = \max_{x_0} \min_{t_0} \left\{ t_0 / \forall t \geq t_0,
\Delta(t,x_0) \leq \varepsilon \right\}.
\end{equation}

In this letter, we prove that for a tiling of $N_T$ tiles
\begin{equation}
\tau(\varepsilon) \leq Cst. \; (N_T)^\nu \ln N_T \ln (1/\varepsilon).
\label{upperbound}
\end{equation}
More precisely, we establish, with the help of reduced numerical work,
that such a bound holds for $\nu=3$, then we argue that $\nu=2$
should be the correct exponent. Whatever this exponent, this proves that
flip dynamics are rapidly mixing at infinite temperature. As for the
$\ln(N_T)$ correction in~(\ref{upperbound}), as discussed
in~\cite{Wilson99}, it is a feature of our choice of distance
$\Delta(t,x_0)$: an Euclidean norm would not display this correction
but it is less natural in the context of measure of probability
distributions convergence.

The coupling technique~\cite{Aldous81} has been successfully applied
to estimate mixing times of several systems, such as hexagonal
tilings~\cite{Randall,Randall98,Wilson99}. It relies on the surprising
idea that following the dynamics of {\em couples} of configurations
instead of a single one might provide the properties of the original
dynamics on single configurations. A {\em coupling} is a Markov chain
on $L \times L$; couples of configurations are updated simultaneously
and are strongly correlated, but each configuration, viewed in
isolation, performs transitions of the original Markov
chain. Moreover, the coupled process is designed so that when both
configurations happen to be identical, then they follow the same
evolution and remain identical forever. Then the central idea of the
technique is that the average time the two configurations need to
couple (or to {\em coalesce}) provides a good upper bound on the
original mixing time $\tau(\varepsilon)$: given an initial couple
$(x_0,y_0)$ at time $t=0$, define the {\em coalescence} time
$T(x_0,y_0)$ as the minimum time that both configurations need to
coalesce, and the {\em coupling} time as
\begin{equation}
T=\max_{(x_0,y_0) \in L \times L} <T(x_0,y_0)>,
\label{cplg_time}
\end{equation}
where the last mean is taken over realizations of the coupled Markov
chain. The following theorem, central in the coupling approach,
provides an upper bound for the ergodic times of the original ({\em
not coupled}) process~\cite{Aldous81}: 
\begin{equation}
\tau(\varepsilon) \; \leq \; T e \ln(1/\varepsilon) + 1 \; \simeq \; T
e \ln(1/\varepsilon),
\label{tau_eps}
\end{equation}
where $e=\exp(1)$. If the configuration set $L$ can be
endowed with a partial order relation $\succeq$ with unique minimum
and maximum elements $\hat{0}$ and $\hat{1}$, the implementation of
the technique is highly facilitated, provided the coupled dynamics is
{\em monotonous}, {\em i.e.} if $x(t) \succeq y(t)$, then $x(t+1) \succeq
y(t+1)$: let $(x_0,y_0)$ be any initial couple such that
$\hat{1} \succeq x_0 \succeq y_0 \succeq \hat{0}$. After any number of
steps, the 4 configurations remain in this order. When the
iterates of $\hat{0}$ and $\hat{1}$ have coalesced, the iterates of
$x_0$ and $y_0$ also have. Thus $T(x_0,y_0) \leq T (\hat{0}, \hat{1})$
and $T = \; <T (\hat{0}, \hat{1})>$.

\begin{figure}[ht]
\begin{center}
\ \psfig{figure=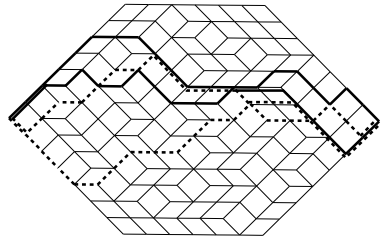,width=5cm} \
\end{center}
\caption{Example of 2-line coupling in the directed path
representation. Dark lines represent a configuration and dotted lines
the second one. Both configurations form a couple.}
\label{cpls.figs}
\end{figure}

A convenient representation of random rhombus tilings was introduced
by de Bruijn~\cite{DeBruijn81}. It consists of following in a tiling
lines made of adjacent tiles sharing an edge with a fixed orientation
(Fig.~\ref{flip2D}). The set of lines associated with an orientation
is called a de Bruijn family. In an octagonal tiling, there are 4
families. When removing a family from an octagonal tiling, one gets an
hexagonal tiling. Conversely, this remark enables one to propose a
convenient construction of tilings~\cite{Mosseri93,Widom98,Bibi99}:
directed paths are chosen on an hexagonal tiling, called the {\em
base} tiling. They are represented by dark lines in
Fig.~\ref{cpls.figs}. They go from left to right without crossing (but
they can have contacts). When they are ``opened'' following a new edge
orientation, they generate de Bruijn lines of the fourth family. In
this one-to-one representation, a tiling flip involving tiles of the
fourth family becomes a path flip: the path jumps from one side of a
tile to the opposite side.

Defining couplings on the whole octagonal tiling sets seems to be an
infeasible task. We instead use the paths-on-tiling point of view to
decompose the configuration space into smaller subsets where couplings
can be defined: let $J_a$ denote the set of tilings which have the
same base hexagonal tiling $a$, called ``fibers''~\cite{Bibi99}. $L$
is a disjoint union of fibers. The only possible flips inside $J_a$
are those which involve the fourth de Bruijn family. Note that 
we can construct four such fibrations, one for each family.

Now, let ${\cal M}$ denote the symmetric transition matrix associated with
the Markov chain on the whole set $L$: given two
configurations $x$ and $y$, the matrix entry ${\cal M}(x,y)$ is equal
to the transition probability $P(x,t+1 | y,t)$.  In the same way, we
define the symmetric transition matrices ${\cal M}_i$ associated with the
Markov chains where only flips involving the $i$-th de Bruijn family
are allowed. Since fibers have been disconnected, ${\cal M}_i$ is
block-diagonal. The following result interconnects the four
fibrations:
\begin{equation}
{\cal M} = ({\cal M}_1 + {\cal M}_2 + {\cal M}_3 + {\cal M}_4)/3 -
\mbox{Id}/3,
\label{zerelation}
\end{equation}
where $\mbox{Id}$ is the identity. Indeed, each coefficient
${\cal M}(x,y)$ appears in all four matrices ${\cal M}_i$ but one, since the
corresponding flip involves 3 de Bruijn lines.

Now we implement the above {\em coupling technique on each fiber}. To
begin with, we suppose that there is only one line in the flipping
family, denoted by $\ell$. As in reference~\cite{Randall}, in order
to have a monotonous coupling, we slightly modify the Markov chain: at
each step, we choose uniformly at random an internal vertex of $\ell$,
the $n$-th one starting form the left, {\em and} a number $r \in
\{ 0,1 \}$. If this vertex is flippable upward (resp. downward) and $r
=0$ (resp. $r=1$), then we flip it. Note that this Markov
chain has a time unit different from the original one.

We now define the order relation ($\succeq$): given two lines
$\ell_1$ and $\ell_2$, $\ell_1 \succeq \ell_2$ if $\ell_1$ is entirely
above $\ell_2$. The maximum (resp. minimum) configuration clearly lies
on the top (resp. bottom) boundary of the hexagonal domain. If the two
flips on $\ell_1$ and $\ell_2$ occur with same $n$ and $r$ and if
$\ell_1 \succeq \ell_2$ then their images satisfy the same
relation. Indeed, as in reference~\cite{Randall}, thanks to the
introduction of $r$, if a flip could bring $\ell_1$ below $\ell_2$,
then the same flip would also apply to $\ell_2$, thus preserving the
order between lines: the coupling is monotonous.

In the general case with $p$ non-intersecting lines in each
configuration (Fig.~\ref{cpls.figs}), let us denote by $\ell_i^{(j)}$,
$j=1,\ldots,p$, the $p$ lines of each configuration $\gamma_i$. Then
$\gamma_1 \succeq \gamma_2$ if for any $j$, $\ell_1^{(j)} \succeq
\ell_2^{(j)}$. The configuration $\gamma$ is maximum (resp. minimum)
when each of its lines is maximum (resp. minimum). At each step, the
index $j$ of the line to be flipped is chosen between 1 and $p$, the
same $j$ for both~$\gamma_i$.

\begin{figure}[ht]
\begin{center}
\ \psfig{figure=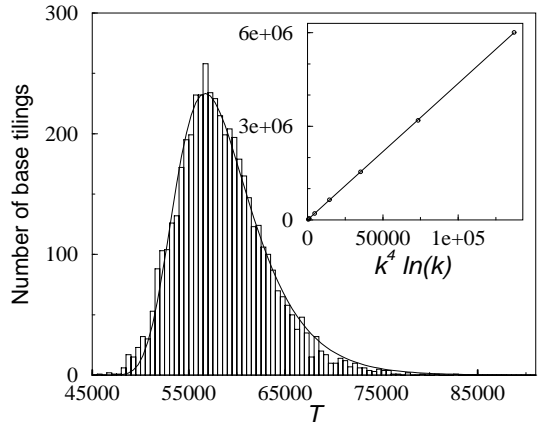,height=5.85cm,width=7cm} \ 
\end{center}
\caption{A distribution of coupling times $T$ in the case
of a diagonal base tiling ($k=10$) and $p=1$, fitted by a Gumbel
distribution (continuous curve; see eq. (\ref{Gumbel})). Inset: in the
diagonal case ($k=p$), numerical estimates of $\overline{T}$ in
function of $k^4 \ln k$ (circles), up to $k=15$, and linear fit. Error
bars are smaller than the size of symbols. The slope is $25.51 \pm
0.05$.}
\label{numerique}
\end{figure}

To begin with, we numerically study the diagonal case, where the 4
sides of the octagonal tilings are equal to $k$.  For a given base
tiling ${\cal T}_a$, we run a number $m$ of couplings until they coalesce,
and then estimate the coupling time $T({\cal T}_a)$. We then make a second
average on $M$ different tilings, in order to get the time
$\overline{T}$ averaged on tilings ${\cal T}_a$. We also keep track of the
standard deviation $\Delta T$. From our numerical data (see
Fig.~\ref{numerique}), we draw the following conclusions: $\Delta T /
\overline{T}$ decreases toward a constant ($\simeq 0.07$) as $k \rightarrow
\infty$, which means that the average coupling time $T({\cal T}_a)$ goes on
depending on the base tiling ${\cal T}_a$ at the large size limit. However,
most $T({\cal T}_a)$ are of order $\overline{T}$, and the mixing times
$\tau(\varepsilon)$ on most fibers are controlled by
$\overline{T}$. Nevertheless, the effect of few ``slower'' fibers will
deserve a detailed discussion below. Moreover, the measures of
$\overline{T}$ are compatible with a $k^4 \ln k$ behavior
(Fig.~\ref{numerique}, inset).  In particular, this fit with
logarithmic corrections is much better than a simple power-law
fit. This result is consistent with known results in the case of
hexagonal tilings~\cite{Randall,Wilson99}, where $T$ also
grows like $k^4 \ln k$.

We also have explored coupling times on fibers in non-diagonal cases
and our conclusions remain identical. As a consequence, couplings in
fibers behave like couplings in hexagonal tiling problems, up
to different numerical prefactors: the dynamics on each fiber is
rapidly mixing. 

\begin{figure}[ht]
\begin{center}
\ \psfig{figure=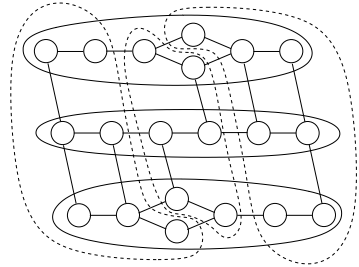,height=3.5cm} \
\end{center}
\caption{The lattice of tilings filling an octagon of sides 1,1,1 and
2. The edges represent possible flips. Two fibrations among four are
represented (continuous and dotted lines). }
\label{fc4_2}
\end{figure}
Now we return to the dynamics on the whole set of tilings. Examining
Fig.~\ref{fc4_2}, one remarks that two fibrations are to a certain
extent ``transverse'': in octagonal tiling sets, one can connect any
two tilings using flips of only two fibers~\cite{Draft}; if the
dynamics is rapidly mixing in each fiber, the combination of dynamics
on two (and even four) fibrations will certainly also be rapidly
mixing. We now establish properly this point: it is common in the
field of Markov processes to relate rates of convergence to spectra of
transition matrices. Generally speaking, given a transition matrix
$M$, 1 is always the largest eigenvalue in modulus, and the difference
$g(M)$ between 1 and the second largest eigenvalue is called the first
gap of $M$. Then $\tau(\varepsilon) \simeq \ln(1/\varepsilon) /
g({\cal M})$ for small $\varepsilon$~\cite{Randall98}. By simple arguments
from linear algebra and Euclidean geometry (since matrices are
symmetric), the following central gap relation can be
established~\cite{Draft}, based upon~(\ref{zerelation}) and the
smallness of the intersection of two fibers:
\begin{equation}
g({\cal M}) \geq \inf_i\left(g({\cal M}_i)\right).
\label{mino}
\end{equation}
This result restores the symmetry lost in the fibration process.  It
implies that the mixing time $\tau(\varepsilon)$ on $L$ is smaller
than the mixing time on the slowest fibration derived
from~(\ref{tau_eps}). Hence $\tau(\varepsilon)$ is smaller than the
mixing time on the slowest fiber. We have seen that the average
coupling time $\overline{T}$ grows like $k^4 \ln k$ in diagonal
cases. Since the number of tiles is $N_T = 6 k^2$, $\overline{T}
\simeq \kappa_4 N_T^2 \ln (N_T)$, where $\kappa_4 = 1.189 \pm 0.003$
in the original time unit. However, the coupling time $T$ depends on
the base tiling ${\cal T}_a$ and the distribution of times
$T$ in a given fibration have a certain width around $\overline{T}$
(Fig.~\ref{numerique}). And even if the typical values of coupling
times are of the order of magnitude of the previous average value,
this does not exclude the existence of rare slow fibers in the upper
tails of these distributions. However, we recall that coupling times
$T$ are maxima of coalescence time distributions
(\ref{cplg_time}). Therefore the expected shape of the distribution of
$T({\cal T}_a)$ ought to be sought in the specific class of
extreme-value distributions, namely Gumbel
distributions~\cite{Gumbel}: consider $N$ independent identical random
variables $T_a$, whose probability densities decay rapidly at large
$T$:
\begin{equation}
p(T) \simeq {C_1 \over T^{\alpha}} \exp ( -C_2 T^{\beta}),
\end{equation}
where $C_1,C_2,\beta>0$. If $T_{\mbox{\scriptsize max}}
= \max_a T_a$, then at large $N$, the probability density of
$T_{\mbox{\scriptsize max}}$ satisfies
\begin{equation}
p(u) = \exp (-u - \exp (-u)),
\label{Gumbel}
\end{equation}
where $u=(T_{\mbox{\scriptsize max}}-T_0)/\delta T$ is a suitably rescaled variable. 

Now, even if coalescence times are not strictly speaking independent
variables~\cite{Draft}, our numerical distributions appear to be well
fitted by this kind of distribution (Fig.~\ref{numerique}). This
result provides the large $T$ behavior of coupling time distributions:
$p(T_{\mbox{\scriptsize max}}) \sim \exp(-u) \sim \exp (-
T_{\mbox{\scriptsize max}}/\delta T)$, and therefore an estimation of
the largest coupling time. Let us focus on diagonal cases: $T_0$ as
well as $\delta T$ behave like $k^4 \ln k$. But for a fibration $i$,
there are $N_i$ base tilings ${\cal T}_a$.  Therefore if $T^*$ is the
largest coupling time on all tilings ${\cal T}_a$, it is estimated by
$N_i \exp (- T^*/\delta T) \approx 1$.  Now $N_i$ grows exponentially
with the number of tiles~\cite{Bibi99}: $\ln N_i \approx Cst_1 \;
k^2$.  Thus
\begin{equation}
T^* \approx Cst_2 \; k^6 \ln k \approx Cst_3 \; N_T^3 \ln N_T
\end{equation}
and $\tau(\varepsilon) \leq Cst_4 \; N_T^3 \ln N_T \ln
(1/\varepsilon)$. However these extreme values should not be relevant:
because of the exponential decay of $p(u)$, slow fibers are rare and
can be bypassed {\em via} rapid ones. More precisely, perturbation
theory arguments suggest that rare slow fibers can be seen as a small
perturbation of an otherwise rapid transition matrix and have a
vanishing influence on its spectral gap~\cite{Draft}: they do not
significantly slow rapid dynamics and the typical coupling time
$\overline{T}$ drives the dynamics on fibers, leading to $\nu=2$
in~(\ref{upperbound}) and $Cst. \approx e \; \kappa_4$.

What does this analysis become in the case of larger symmetry tilings
or of higher dimensional tilings, such as icosahedral ones? Plane
rhombus tilings with $2D$-fold symmetry could be addressed by our approach
without significant technical complication, leading to laws similar
to~(\ref{upperbound}), up to different prefactors $Cst.$~\cite{Draft}. As for higher dimensional tilings, the
fibration process remains valid, but the connectivity of fibers is not
established~\cite{Bibi99}, making impossible a na\"{\i}ve
generalization of this approach.

I would like to thank M. Latapy, K. Frahm, R. Mosseri 
and D.S. Dean for fruitful discussions.

\end{document}